\documentclass[12pt]{article}
\usepackage{amsmath,bm}
\usepackage{amssymb}
\usepackage{graphicx}
\usepackage{amsfonts}         
\usepackage{fancybox}   
\usepackage{yfonts}

\renewcommand{\baselinestretch}{1.4}
\def\comments#1{}

\def\vev#1{\langle{#1}\rangle}

\def\CA{{\cal A}}

\def\CE{{\cal E}}
\def\CF{{\cal F}}

\def\CO{{\cal O}}
\def\CP{{\cal P}}

\def\CS{{\cal S}}

\def\II{\relax{I\kern-.10em I}}

\def\IB{\relax{\rm I\kern-.18em B}}
\def\ID{\relax{\rm I\kern-.18em D}}
\def\IE{\relax{\rm I\kern-.18em E}}
\def\IF{\relax{\rm I\kern-.18em F}}
\def\IG{\relax\hbox{$\inbar\kern-.3em{\rm G}$}}
\def\IGa{\relax\hbox{${\rm I}\kern-.18em\Gamma$}}
\def\II{\relax{\rm I\kern-.18em I}}
\def\IK{\relax{\rm I\kern-.18em K}}

\def\inbar{\,\vrule height1.5ex width.4pt depth0pt}

\def\frac#1#2{{#1 \over #2}}

\newdimen\tableauside\tableauside=1.0ex
\newdimen\tableaurule\tableaurule=0.4pt
\newdimen\tableaustep
\def\phantomhrule#1{\hbox{\vbox to0pt{\hrule height\tableaurule width#1\vss}}}
\def\phantomvrule#1{\vbox{\hbox to0pt{\vrule width\tableaurule height#1\hss}}}
\def\sqr{\vbox{%
  \phantomhrule\tableaustep
  \hbox{\phantomvrule\tableaustep\kern\tableaustep\phantomvrule\tableaustep}%
  \hbox{\vbox{\phantomhrule\tableauside}\kern-\tableaurule}}}
\def\squares#1{\hbox{\count0=#1\noindent\loop\sqr
  \advance\count0 by-1 \ifnum\count0>0\repeat}}
\def\tableau#1{\vcenter{\offinterlineskip
  \tableaustep=\tableauside\advance\tableaustep by-\tableaurule
  \kern\normallineskip\hbox
    {\kern\normallineskip\vbox
      {\gettableau#1 0 }%
     \kern\normallineskip\kern\tableaurule}%
  \kern\normallineskip\kern\tableaurule}}
\def\gettableau#1 {\ifnum#1=0\let\next=\null\else
  \squares{#1}\let\next=\gettableau\fi\next}

 \def\eqnn#1{\xdef #1{(\secsym\the\meqno)}\writedef{#1\leftbracket#1}%
 \global\advance\meqno by1\wrlabeL#1}
 \def\eqna#1{\xdef #1##1{\hbox{$(\secsym\the\meqno##1)$}}
 \writedef{#1\numbersign1\leftbracket#1{\numbersign1}}%
 \global\advance\meqno by1\wrlabeL{#1$\{\}$}}
 \def\eqn#1#2{\xdef #1{(\secsym\the\meqno)}\writedef{#1\leftbracket#1}%
 \global\advance\meqno by1$$#2\eqno#1\eqlabeL#1$$}

\global\newcount\itemno \global\itemno=0

\def\itemaut#1{\global\advance\itemno by1\noindent\item{\the\itemno.}#1}

\def\del{\partial}

\def\({\left(}
\def\){\right)}

\def\eg{{\it e.g.}}
\def\ie{{\it i.e.}}

\def\etal{{\it et.\ al.}}
\newif{\ifeq}           
\eqtrue                 
\newcommand{\be}{\begin{equation}}
\newcommand{\ee}{\end{equation}}
\newcommand{\bea}{\begin{eqnarray}}
\newcommand{\eea}{\end{eqnarray}}
\newcommand{\bean}{\begin{eqnarray*}}
\newcommand{\eean}{\end{eqnarray*}}

\def\({\left(}
\def\){\right)}
\def\[{\left[}
\def\]{\right]}

\newcommand{\half}{\frac{1}{2}}

\renewcommand{\O}{{\cal O}}

\def\CO{\O}

\def\ie{{\it i.e.}}

\newcommand{\lsim}{\,\raise.3ex\hbox{$<$\kern-.75em\lower1ex\hbox{$\sim$}}\,}
\newcommand{\gsim}{\,\raise.3ex\hbox{$>$\kern-.75em\lower1ex\hbox{$\sim$}}\,}

\newcommand{\scz}{\setcounter{equation}{0}}

\newif{\ifeq}
\eqtrue

\numberwithin{equation}{section}

\begin{document}

\begin{titlepage}

\begin{flushright}
MIT-CTP/4068, NSF-KITP-09-136
\end{flushright}
\vfil

\begin{center}
{\huge An analytic Lifshitz black hole}\\
\end{center}
\vfil
\begin{center}
{\large Koushik Balasubramanian and John McGreevy}\\
\vspace{1mm}
Center for Theoretical Physics, MIT,
Cambridge, Massachusetts 02139, USA\\
\vspace{3mm}
\end{center}

\vfil

\begin{center}

{\large Abstract}
\end{center}

\noindent
A Lifshitz point is described by a quantum field theory with anisotropic scale invariance
(but not Galilean invariance).
In 0808.1725,  gravity duals were conjectured for such theories.
We construct analytically a black hole which asymptotes
to a vacuum Lifshitz solution;
this black hole solves the equations of motion
of some simple (but somewhat strange) extensions of the models of 0808.1725.
We study its thermodynamics and scalar response functions.
The scalar wave equation turns out to be exactly solvable.  
Interestingly, the Green's functions do not exhibit the ultralocal
behavior seen previously in the free Lifshitz scalar theory.

\vfill
\begin{flushleft}
August 2009
\end{flushleft}
\vfil
\end{titlepage}
\newpage
\renewcommand{\baselinestretch}{1.1}  

\renewcommand{\arraystretch}{1.5}

\section{Introduction}\scz

\hspace{0.2 in}
A great deal of progress has been made in the study of 
quantum field theories 
and their holographic duals.
The possible scope of this enterprise is not yet clear;
for example, the correspondence seems to extend
to some systems without Lorentz invariance.
Recently, attempts have been made to apply 
the holographic principle to study 
condensed matter systems near a critical point (for reviews, see \cite{Sachdev:2008ba, 
Hartnoll:2009sz}).
There are many scale-invariant field theories that are not Lorentz 
invariant, which are of interest in studying such critical points. 
In such a theory, time and space 
can scale differently 
\ie,
$t \rightarrow \lambda^z t ,\, \vec{x} \rightarrow \lambda \vec{x}$ 
under dilatation. The relative scale dimension of time and space, $z$, is called the `dynamical exponent'.
Such a scale invariance is exhibited by a Lifshitz theory, 
which we will take to mean
an anisotropic scale-invariant theory which is not Galilean-invariant. 
The following Gaussian action provides a simple example
of a (free) Lifshitz theory in $d$ space dimensions:
\be\label{freelifshitzscalar}
 S[\chi]  = \int d^dx dt \[ \( \del_t \chi\)^2 - K \(\nabla^2 \chi\)^2\] \ee
This action describes a fixed line parametrized by $K$, and the dynamical exponent is
$z=2$. This theory describes the critical behavior of \eg\ quantum dimer models \cite{Ardonne:2003wa}.
In many ways, 
the $d=2, z=2$ version of the theory (\ref{freelifshitzscalar}) is like a relativistic boson in 1+1 
dimensions\footnote{Similar statements apply whenever $z=d$.  However, constructing a rotation-invariant, local
spatial kinetic operator that scales like $p^{2d}$ is tricky for $d\neq 2^k$ for integer $k$.
We note in passing that the existence of such theories seem to be 
suggested by the calculations of \cite{Park:2008ki}.}.
The scaling behavior of the ground-state entanglement entropy
for this class of theories was studied recently in \cite{Hsu, Fradkin}.
This analysis also supports the similarity with 2d CFT,
in that a universal leading singular behavior is found.

In the free theory, the boson has logarithmic correlators 
\be
 \vev{\chi(x)\chi(0)} \sim \int d\omega d^2 k {1\over \omega^2 - k^4 } e^{ i \vec k 
 \cdot \vec x - i \omega t } \sim \ln x. 
 \ee
As in the familiar $d=z=1$ case, the operators of definite scaling dimension are not the 
 canonical 
 bose field itself, but rather its exponentials and derivatives.
 In the connection with quantum dimer models, 
 the bose field is a height variable constructed from the dimer configuration,
 and the exponentials
 of the bose field are order parameters for various dimer-solid orderings
 \cite{Ardonne:2003wa}.
At zero temperature, the logarithmic behavior of the correlator of the bose fields implies that the two-point function of the order parameter 
decays as a power law. However, the equal-time correlators at finite temperature 
are {\it ultra-local} in the infinite-volume limit \cite{Senthil}: they 
vanish at any nonzero spatial separation. 
In \cite{Senthil}, it was suggested that this might be a mechanism
for the kind of local criticality (scaling in frequency, but not momentum) 
seen in the strange metal phase of the cuprates and in heavy fermion materials.
One is led to wonder whether 
this property should is shared by interacting Lifshitz theories,
and whether the Lifshitz scaling is sufficient to produce this behavior.
In \cite{Senthil} the addition of perturbative interactions 
was shown to lead to a finite correlation length;
these perturbations violate the Lifshitz scaling. 
Below we will show that 
interactions which preserve the Lifshitz scaling 
need not give ultralocal behavior.
%

Gravity solutions with Lifshitz-type scale invariance were found in \cite {KLM}.
They found that the following family of metrics, parametrized by $z$, provide a geometrical description of Lifshitz-like theories (with $z$ as the dynamical exponent):
\be ds^2 = L^2 \(-{dt^2\over r^{2z}} + {\vec{dx}^2 + dr^2 \over r^2}\), \ee
where $\vec{x}$ denotes a $d-$dimensional spatial vector. 
For $d=2$, this metric extremizes the folowing action:
\be\label{KLMaction} S = \half \int d^4 x \(R - 2 \Lambda\) - \half \int \( F_{(2)} \wedge \star F_{(2)} + F_{(3)} \wedge \star F_{(3)}\) - c \int B_{(2)}\wedge F_{(2)},\ee
where $ F_{(2)}= dA_{(1)}$,  $ F_{(3)}= dB_{(2)}$ and $\Lambda$ is the four dimensional cosmological constant. They computed the two-point function for the case when $z=2$ and showed that it exhibits power law decay. They also studied the holographic renormalization group flow for this case and found that $AdS_4$ is the only other fixed point  of the flow.
Lifshitz vacuum solutions were shown to be stable under perturbations of the 
bulk action in \cite{Adams:2008zk}.
%
%
%
%
%
%
%

In this paper we shall study a black hole solution which asymptotes to the Lifshitz spacetime
with $d=2, z=2$.
In section 2, an analytical solution for a black hole that asymptotes to the planar Lifshitz spacetime is written down.
We present several actions whose equations of motion it solves; they all involve some
matter sector additional to (\ref{KLMaction}).
Section 3 presents an analysis of the thermodynamics of this black hole. In section 4, we 
solve the wave equation for a massive scalar field in this background;
surprisingly, this equation is exactly solvable.
We use this solution 
to calculate the two-point functions of boundary operators in section 5.

Since we found the solution described in this paper, some related work has appeared.
\cite{Taylor:2008tg} constructs a black hole solution in a related background
with slightly different asymptotics.  
Danielsson and Thorlacius \cite{Danielsson} found numerical solutions of 
black holes in global Lifshitz spacetime. 
Interestingly, these are solutions to precisely the system studied by \cite{KLM}, with no additional fields. 
Related solutions were found by \cite{Omid, Bertoldi:2009vn}.
\cite{Azeyanagi:2009pr} found solutions of type IIB supergravity that are dual to Lifshitz-like theories with spatial anisotropy and $z=3/2$; these solutions have a scalar field which breaks the scaling symmetry.
To our knowledge, a string embedding of $z=2$ Lifshitz spacetime is still not known;
obstacles to finding such an embedding are described in \cite{taknogo}.

\section{Black hole solution} 

\subsection{Vacuum solution}

 The tensor fields in \cite{KLM} can be rewritten as one massive gauge field.
The Chern-Simons-like coupling is responsible for the mass.
A familiar example is that of a 2-form field strength $F$ and a 3-form field strength $H$ in five dimensions
with $ L = F \wedge \star F + H \wedge \star H + F \wedge H $:
this gives the same equation of motion as $ L = F \wedge \star F + A^2 $.
In the four dimensional case studied in \cite{KLM}, 
the dual of the 3-form field strength in four dimensions is a scalar field $\varphi$.
Then
\be B_2 \wedge F_2 = - F_3 \wedge A_1 + {\rm bdy~terms} = - \star d \varphi \wedge A_1 = - \sqrt g \del^\mu \varphi A_\mu .\ee
The action then reduces to
\be F_2 \wedge \star F_2 + ( \del \varphi + A )^2 ,\ee
and $\varphi$ shifts under the $A$ gauge symmetry,
and we can fix it to zero, and this is just a massive gauge field\footnote{This
was also observed in \cite{Taylor:2008tg}.}.
Hence,
the zero-temperature Lifshitz metric 
\be ds^2 = -  {dt^2\over r^{2z}} + { d\vec x ^2 + dr^2  \over r^2} ,\ee
is a solution of gravity  in the presence of cosmological constant and a massive gauge field,
and the gauge field mass is $m^2 = d z $.
The bulk curvature radius has been set to one here and throughout the paper;
in these units, the cosmological constant is 
$\Lambda = -{z^2 + (d-1)z + d^2 \over 2}$.
The gauge field profile is $A = \Omega r^{-z} dt$ (in the $r$ coordinate with the boundary at $r=0$),
and the strength of the gauge field is (for $d=2$)
$$ \Omega^2 = {8 { z^2 + z - 2 \over z (z+2)} } . $$

We note in passing that the Schr\"odinger spacetime is a solution of the same action with a different mass for the gauge field and a different cosmological constant \cite{Son:2008ye, Balasubramanian:2008dm}. 
Therefore we find the perhaps-unfamiliar situation 
where the same gravitational action has solutions with very different asymptopia.
Another recent example where this happens is `chiral gravity' in three dimensions,
which has asymptotically AdS solutions as well as various squashed and smushed and wipfed solutions \cite{Li:2008dq}.

Given this fact, one might expect that the Lifshitz spacetime can be embedded into the same type IIB truncations as the Schr\"odinger spacetime (see 
\cite{Herzog:2008wg, Adams:2008wt} and especially 
\cite{Maldacena:2008wh}). However, the scalar equation of motion is not satisfied by 
the Lifshitz background since $F^2$ is non-zero. 

 \subsection{Black hole solution}

We shall now study a black hole in four dimensions that asymptotically approaches the Lifshitz spacetime with $z=2$.  
We first observe that there is such a black hole in a 
 system with a strongly-coupled scalar
(\ie\ a scalar without kinetic terms).  The action is  
\be \label{scalaraction} S_1 = \half\int d^4 x \( R - 2 \Lambda\) - \int d^4 x\({e^{-2\Phi}\over 4}F^2  + {m^2\over 2} A^2 +  \(e^{-2 \Phi} - 1\) \) .
\ee 
A solution of this system is 
$$\Phi = -\half \log \(1+r^2/r_H^2\), ~~~A= f/r^2 dt $$
 \be \label{LifBHmetric}ds^2 = - f {dt^2\over r^{2z}} + { d\vec x ^2 \over r^2} + {dr^2  \over f r^2}, 
\ee
with 
$$ f = 1-{r^2\over r_H^2}. $$
Note  that the metric has the same simple form as in the RG flow solution (eqn (4.1)) of \cite {KLM}.

We can get the same contributions to the stress tensor
as from the scalar without kinetic terms from several more-reasonable systems.
One such system is obtained by adding a second massive gauge field $B$
which will provide the same stress-energy as the scalar.  It has a slightly unfamiliar action:
\be \label{twogaugefieldsaction} 
S_2  = \half \int d^4 x \(R - 2\Lambda\) - \int d^4 x \({1 \over 4}B^2 dA^2 + {m_A^2\over 2} A^2 + {1\over 4}dB^2 - {m_B^2 \over 2}(1- B^2)\) \ee
where $A,B$ are one-forms, and $m^{2}_{A}=4$ and $m^{2}_{B} = 2$ .
The solution looks like
$B = B(r)dr,  A = A(r) dt $
and the metric is same as (\ref {LifBHmetric}).
In the solution, the scalar functions take the form 
$$ 
B(r) = \sqrt{g_{rr} \(1 + {r^2\over r_H^2}\)}, ~~A(r) =  \Omega f r^{-z} dt  $$
Note that $B(r)$ isn't gauge-trivial (even though its field strength vanishes) because of the mass term. 
Since $B(r)$ asymptotes to $1$, the effective gauge coupling of the field $A$ is not large at the boundary.

The system with a strongly-coupled scalar in (\ref {scalaraction}) is not equivalent to the system 
(\ref{twogaugefieldsaction}) with two gauge fields. For example, 
there are solutions of (\ref {scalaraction}) where the scalar has a profile that depends both on $r$ and $x$;
such configurations do not correspond to solutions of (\ref{twogaugefieldsaction}).

It is not clear whether the solution written above is stable. We leave the analysis of the stability of such solutions 
to small perturbations to future work.
As weak evidence for this stability, we show in the next section that these 
black holes are {\it thermodynamically} stable.


Another action with this Lifshitz black hole 
(\ref{LifBHmetric})
as a solution is
\be\label{actiontwo}
S_3 = \half \int d^4 x \( R- 2 \Lambda - \half dB^2 - (\del \Phi - B)^2 - m_AA^2 - \half  e^{- 2 \Phi} F^2 - V(\Phi) \)
\ee
where $V(\Phi) =   2 e^{ - 2\Phi} - 2  $. 
In the solution, the metric and gauge field $A$ take the same form as in (\ref{LifBHmetric}).
The other fields are
$$ e^{-2 \Phi}  =  1 + {r^2 \over r_H^2} , ~~~~~B = d \Phi .$$

Note that the action (\ref{actiontwo}) is {\it not} invariant under the would-be gauge transformation 
$$ B \to B + d\Lambda, ~~~~~\Phi \to \Phi + \Lambda,$$
because of the coupling to $F^2 - 4$ 
(the sum of the gauge kinetic term and the potential term)\footnote{
We note that 
this quantity does vanish on the solution of interest.}.
We are not bothered by this: it means that in quantizing the model,
mass terms for the fluctuations $B$ will be generated; however, 
such a mass term is already present.

We would also like to point out that 
in the three systems $S_{1,2,3}$ described above,
the stress-energy tensor of the fields with local propagating degrees of freedom satisfy the dominant energy condition\footnote{We would like to thank Allan Adams, Alex Maloney and Omid Saremi for bringing this criterion to our attention.}, 
\ie\  $~T_{\mu \nu}^{(\Phi,A,B)} \( = R_{\mu\nu} - (\half R + \Lambda) g_{\mu \nu}\)$ satisfies the following 
$${T_{tt} \over T_{xx}} = {T_{tt} \over T_{yy}} > -1 \text{ and } {T_{tt} \over T_{rr}} > -1.$$
Hence, there are no superluminal effects in the bulk.
This is basically a consequence of the fact that the squared-masses of the gauge fields are positive. 

\section{Lifshitz black hole thermodynamics}

The Hawking temperature and entropy can be calculated using the near horizon geometry. 
The Hawking temperature is the periodicity of the Euclidean time direction in the near horizon metric (proportional to the surface gravity) \ie, $T = {\kappa\over 2 \pi} |_{r=r_H} $, with 
$$ \kappa^2 = - \half \nabla^a v^b \nabla_b v^a $$ 
where $ v = \del_t $. Hence,
\be \label{tempBH} T = {1\over 2 \pi r_H^2} .\ee
The entropy of the black hole is  
\be {\CS} = {\text{ Area of Horizon}\over 4 G_N}  = { L_x L_y \over 4 G_4 r_H^2}. \ee
\hspace{0.2 in}We shall now evaluate the free energy, internal energy and pressure by calculating the on-shell action and boundary stress tensor. In order to renormalize 
the action, it is essential to add counterterms which are intrinsic invariants of the boundary (see \cite{VBPK}). 
 
Consider the following gravitational action:
$$ S  =  \half \int_M d^4 x \sqrt{g} \( R - 2 \Lambda -
{e^{-2\Phi} \over 4} F^2 - {m_A^2\over 2} A^2 -  V(\Phi)\) $$ 
\be \label{onshell} - \int_{\del M} d^3x\sqrt {\gamma} \( K + {c_N}  e^{-2 \Phi}n^{\mu}A^{\nu}F_{\mu \nu}  \) 
\ee
 $$
+ ~\half \int_{\del M}d^3x \sqrt{\gamma} \( 2c_0 -  c_1 \Phi -  c_2 \Phi^2\)
+ \half \int_{\del M} d^3 x \sqrt{\gamma} \( \(c_3 + c_4 \Phi \) A^2 + c_5 A^4\) ~.$$ 
The second line of (\ref{onshell}) contains extrinsic boundary terms: the 
Gibbons-Hawking term, 
and a `Neumannizing term' 
which changes the boundary conditions on the gauge field.
The last line of (\ref{onshell}) describes the intrinsic boundary counterterms\footnote{The most general combination of counterterms, which do not vanish at the boundary, is 
$$\half \int_{\del M}d^3x \sqrt{\gamma} \( 2c'_0 +  c'_1 \Phi +  c'_2 \Phi^2\)  + 
\half \int_{\del M} d^3 x \sqrt{\gamma} \( \(c'_3 + c'_4 \Phi \) (A^2-1) + c'_5  (A^2-1)^2\)$$which has the same form as (\ref{onshell}).
}.
In the above expression, we have set $8 \pi G = 1$. 
We have written the analysis in terms of $S_1$ (\ref{scalaraction});
the analysis can be adapted for $S_2$ (\ref{twogaugefieldsaction})
by simply replacing $\Phi$ in (\ref{onshell}) by $-\half \log B^2$.
If Neumann boundary conditions are imposed on the gauge field, then $c_N = 1$ and $c_i = 0$ for $i\ge 3$. Similarly, $c_N=0$, if Dirichlet boundary condition is imposed on the gauge field. 

The boundary stress tensor resulting from (\ref{onshell}) is
$$ T_{\mu \nu} = K_{\mu \nu} - \(K - c_0 +\half c_1 \Phi + \half c_2 \Phi^2\)  \gamma_{\mu \nu} + {e^{-2\Phi}\over 2}\(n^r A_{\mu} \del_r A_{\nu} + n^r A_{\nu} \del_r A_{\mu} - n^r A_{\alpha}\del_rA^{\alpha}\gamma_{\mu\nu}\) $$
\be+\(c_3 + c_4 \Phi + 2 c_5 A^2\)A_\mu A_\nu - \half \(c_3 + c_4 \Phi +  c_5 A^2\)A^2
\gamma_{\mu\nu}
\ee

The values of $c_i$ are determined by demanding that the action is ``well-behaved". The action is well-behaved if the variation of the action vanishes on-shell and if the residual gauge symmetries of the metric are not broken.
The values of $c_i$ which makes the action well-defined also render finite the action and boundary stress tensor (please see appendix A).
Implementing this procedure, we find 
for the energy density, pressure and free energy
\be \label{mainEPF} \CE = \CP = - \CF = \half T \CS = {L_{x} L_{y} \over 2 r_{H}^{4}}\ee
Satisfying the first law of thermodynamics (in the Gibbs-Duhem form $\CE + \CP = T\CS$)
is a nice check on the sensibility of our solution,
since it is a relation between near-horizon ($T, \CS$) and near-boundary ($\CE, \CP$) quantities.

Recently, \cite{Ross:2009ar} have described an alternative set of 
boundary terms for asymptotically Lifshitz theories.  
They do not include the Neumannizing term, but instead include an
intrinsic but nonanalytic $\sqrt{A^\mu A_\mu}$ term.

\section{Scalar response}

In this section, we study a probe scalar in the black hole background (\ref{LifBHmetric}).
The scalar can be considered a proxy for the mode of the metric coupling to $T^x_y$.

 \subsection{Exact solution of scalar wave equation}

 Consider a scalar field $\phi$ of mass $m$ in the black hole background 
(\ref{LifBHmetric})\footnote{In the following we have set both the bulk radius of curvature 
and the horizon radius to one.
 This means that frequencies and momenta are `gothic' \cite{Son:2002sd}, \ie\ measured in units of $r_H$.  
 Note that since $z=2$, $\omega$ needs two factors of $r_H$ to make
 a dimensionless quantity.}.

 Let $u \equiv {r^2\over r_H^2}$.
Fourier expand:
 $$ \phi = \sum_k \phi_k(u) e^{ - i \omega t + i \vec k \cdot \vec x } $$
 The wave equation takes the form:
 $$
0 = \frac{u \left(-f k^2+u \omega ^2\right)+m^2 f  }{4 f^2 u^2} \phi_k(u) -\frac{1}{f u}\phi_k'(u)+\phi_k''(u)
 $$
 where $ k^2 \equiv \vec k^2 $ .
Near the horizon, the incoming ($-$) and outgoing ($+$) waves are 
$$ \phi_k \sim ( 1 - u ) ^{ \pm i \omega/2} .$$
The solutions near the boundary at $u=0$ are 
$$ \phi_k \sim u^{ 1 \pm \half \sqrt{ 4 + m^2 } }$$
The {\it exact} solution to the wave equation is
$ \phi_k(u) = f^{ - i \omega/2} u^{1 - \half \sqrt{m^2+4}} G_k(u) $ with 
\begin{equation}
\label{gerbil}
 G_k(u) = 
A_1 ~{}_2F_1 (a_+,b_+;c_+, u)u^{\sqrt{m^2+4}} + 
   A_2 ~ {}_2F_1  (a_-, b_-;c_- , u) \end{equation}
and    
\begin{eqnarray*}
&(a_\pm,b_\pm;c_\pm)\equiv \qquad\qquad \cr
&\left(-\frac{i \omega }{2}\pm \frac{\sqrt{m^2+4}}{2}-\frac{1}{2} \sqrt{-k^2-\omega ^2+1}+\frac{1}{2}, 
-\frac{i \omega}{2}\pm \frac{\sqrt{m^2+4}}{2}+\frac{1}{2} \sqrt{-k^2-\omega ^2+1}+\frac{1}{2};
1 \pm\sqrt{m^2+4};u\right)  
\end{eqnarray*}
   
We emphasize that this is the exact solution to the scalar wave equation in this black hole;
such a solution is unavailable for the AdS$_{d>3}$ black hole.  The difference 
is that the equation here has only three regular singular points, whereas
the AdS$_5$ black hole wave equation has four.  This is because in the AdS$_5$ black hole, the emblackening
factor is $f = 1 - u^2 $ which has two roots, whereas ours is just $ f = 1 - u$.  

The other example of a black hole with a solvable scalar wave equation 
is the BTZ black hole in $AdS_3$ \cite{Birmingham:2001dt}\footnote{
Another example, in two dimensions, is \cite{Peet:1993vb}.}.  The origin of the solvability in that case is the 
fact that BTZ is an orbifold of the zero-temperature solution.
This is {\it not} the origin of the solvability in our case --
this black hole is not an orbifold of the zero-temperature solution.  
This may be seen by comparing curvature invariants: they are not locally 
diffeomorphic. 
More simply, if the black hole were an orbifold, 
it would solve the same equations of motion as the vacuum solution.
The fact that we were forced to add an additional matter sector (such as $\Phi$ or $B_\mu$)
to find the black hole solution immediately shows that 
they are not locally diffeomorphic.
 
Now we ask for the linear combination of (\ref{gerbil}) which is ingoing at the horizon.
In terms of $ \nu \equiv \sqrt{4 + m^2}$, $\gamma \equiv \sqrt{ 1 - \omega^2 - k^2 } $, this is 
the combination with
\be\label{aovera} 
 {A_1 \over A_2} = 
 -  (-1)^\nu {\Gamma (\nu) \over \Gamma(-\nu) } 
 { \Gamma\left ( \half( 1 - i \omega - \nu - \gamma) \right) 
 \over \Gamma\left ( \half( 1 - i \omega + \nu - \gamma) \right) }
  { \Gamma\left ( \half( 1 - i \omega - \nu + \gamma) \right) 
 \over \Gamma\left ( \half(1  - i \omega + \nu+ \gamma) \right) } .
 \ee
 
In the massless case, one of the hypergeometric functions in (\ref{gerbil})
specializes to a Meijer G-function, and the solution is 
$ \phi_k = u^2 f^{- i \omega/2} G_k(u) $ with 
\begin{eqnarray*}
& G_k(u) = \cr
& c_2 \, _2F_1\left(-\frac{i \omega }{2}-\frac{1}{2} \sqrt{-k^2-\omega ^2+1}+\frac{3}{2},-\frac{i \omega }{2}+\frac{1}{2}
   \sqrt{-k^2-\omega ^2+1}+\frac{3}{2};3;u\right)
   + \cr \cr
   &c_1 G_{2,2}^{2,0}\left(u\left|
\begin{array}{c}
 \frac{1}{2} \left(i \omega -\sqrt{-k^2-\omega ^2+1}-1\right),\frac{1}{2} \left(i \omega +\sqrt{-k^2-\omega ^2+1}-1\right) \\
 -2,0
\end{array}
\right.\right)
\end{eqnarray*}
In this solution, the coefficient of $c_1$ (the Meier-$G$ function) is purely ingoing at the horizon.

\subsection{Correlators of scalar operators}

In the previous section we wrote the solution for the wave equation in this black hole
for a scalar field with an arbitrary mass. As mentioned earlier, the BTZ black hole also shares this property of having a scalar wave equation
whose solutions are hypergeometric. Hence, one might expect that the two-point
function of scalar operators in a Lifshitz-like theory to have a form that is similar to that of 2D CFTs.

The momentum space correlator 
for a scalar operator of dimension $\Delta = \Delta_-$
is determined from the ratio of the non-normalizable and normalizable parts of the solution. The asymptotic behavior of the solution in (\ref{gerbil}) is 
\be \phi \sim  u^{{\Delta_{+} \over 2}} \(A_{1} + \CO(u) \) + u^{{\Delta_{-} \over 2}} \(A_{2} + \CO(u) \) \ee 
Hence, the retarded Green's function (two-point function) is 
\be G_{\text{ret}}(\omega, \vec{k}) = -{A_{1}\over A_{2}} = (-1)^\nu {\Gamma (\nu) \over \Gamma(-\nu) } 
 { \Gamma\left ( \half( 1 - i \omega - \nu - \gamma) \right) 
 \over \Gamma\left ( \half( 1 - i \omega + \nu - \gamma) \right) }
  { \Gamma\left ( \half( 1 - i \omega - \nu + \gamma) \right) 
 \over \Gamma\left ( \half(1  - i \omega + \nu+ \gamma) \right) } 
\ee
with $\nu$ and $\gamma$ defined above equation (\ref{aovera}).
Note that the correlator has a form very similar to that of a 2D CFT. It would be nice to know the precise connection between $z=2$ Lifshitz-like theories in $2 + 1$ D with 2D CFTs that
is responsible for this similarity. Note that the poles of the retarded Green's function do not lie on a straight line in the complex frequency plane, as they do for 2D CFTs. 

Next, we would like to see whether the correlators exhibit ultra local behavior 
at finite temperature as observed in the free scalar Lifshitz theory \cite {Senthil}. 
We find that the Green's function is not ultra-local -- 
this removes the possibility that Lifshitz-symmetric interactions 
require ultralocal behavior.

We will now calculate the two-point function of a scalar operator of dimension  $\Delta = 4$ at finite temperature. In this case, the correlator is given by the coefficient of $r^4$ in the asymptotic expansion of the solution near $r=0$. Kachru \etal\ \cite{KLM} showed that the correlator exhibits a power law decay at zero temperature.

We can evaluate this 
correlator by extracting the coefficient of the $u^2$ term (note that $u \propto r^2$) in the asymptotic expansion of the solution of the massless scalar wave equation. The behavior of the solution near $u=0$ is 
$$ \phi (u, \vec{k}, \omega) = 1 - {u\over 4} \( \vec{k}^2 + 2 i \omega\) - {u^2 \over 64}\( \( \vec{k}^2\)^2 + 4 \omega^2\) \Bigg[-3 + 2 \psi\(\half\(-1 + i \omega - \sqrt{1-\vec{k}^2 - \omega^2}\) \) $$ \be+ 2 \psi\(\half\(-1 + i \omega + \sqrt{1-\vec{k}^2 - \omega^2}\) \) + 2\gamma_E+ 2 \ln u  \Bigg] + {\cal O}\(u^3\)\ee
where $\gamma_E$ is Euler's constant, $\psi$ is the digamma function. The behavior of the solution in the Euclidean black hole can be obtained by replacing $\omega$ by $-i |\omega| $. The choice of the negative sign gives the solution which is ingoing at the horizon, as appropriate to the retarded correlator \cite{Son:2002sd}. Henceforth, we shall work with the solution for the Euclidean case. The correlator is the sum of the two digamma functions. All other terms in the coefficient of $u^2$ are contact terms. Hence, the correlator in momentum space is 
$$ \vev{\CO{(-\omega,-\vec{k})}\CO{(\omega,\vec{k})}}  \propto  \( \( \vec{k}^2\)^2 - 4 \omega^2\) \Bigg[\psi\(\half\(-1 + | \omega| - \sqrt{1-\vec{k}^2 + \omega^2}\) \) + $$ \be \psi\(\half\(-1 + | \omega| + \sqrt{1-\vec{k}^2 + \omega^2}\) \)  \Bigg]\ee
After dropping the contact terms, the above expression can be written as follows
$$ \vev{\CO{(-\omega,-\vec{k})}\CO{(\omega,\vec{k})}}  \propto   \sum_{n=1}^{\infty} {\CA_n}$$
 where
 $$\CA_n = { (2n - 3)  + |\omega| \over (2n - 3)^2 + 2 |\omega|(2n - 3) + k^2 - 1 } = {a_n + \mid  \omega \mid \over a_n^2 + 2a_n \mid \omega \mid + k^2 - 1} $$
We can now calculate the correlators in coordinate space by performing the Fourier transform 
of the above expression\footnote {We would like to thank Shamit Kachru and Mike Mulligan for sharing the Mathematica file that computes the spatial two-point function derived in \cite{KLM}.}.  This is given by
\be \label{twopt} D(|\vec{x}|,t) =  \Big[(4\del^2_t - (\nabla^2)^2) \Big]\sum_n \CF_n \ee
where, $D(|\vec{x}|,t)$ is the two-point function and 
 $$ \sum_n \CF_n = \sum_n\int k dk d\omega d\theta {a_n + \mid  \omega \mid \over a_n^2 + 2a_n \mid \omega \mid + k^2 - 1/4}e^{ik|\vec{x}|\cos\theta + i \omega t} $$ 
The short distance ($r \ll r_H$) behavior of the equal time correlator is 
 \be D(|\vec{x}| \ll r_H,0) = \Big[(4\del^2_t - (\nabla^2)^2) \CF \Big]_{t=0,|\vec{x}|\rightarrow0} \propto{1 \over |\vec{x}|^8} .\ee
  As a check, 
 we note that, the short distance behavior 
 of this expression reproduces the zero-temperature answer $|\vec x|^{-8}$ found in \cite{KLM}.

The long distance ($|\vec{x}| \gg r_H$) behavior is
 \be D(|\vec{x}| \gg r_H,0) = \Big[(4\del^2_t - (\nabla^2)^2) \CF \Big]_{t=0,|\vec{x}|\rightarrow \infty} \propto {e^{-\sqrt{2}|\vec{x}|/r_H}\over |\vec{x}|^{3/2}} .\ee
The correlator is not ultra-local, unlike the thermal correlator in free scalar Lifshitz theory.

\section{Outlook}

An important defect of our work which cannot have avoided the reader's attention
is the fact that the matter content which produces the stress-energy tensor for this black hole is unfamiliar and
contrived. 
There is no physical reason why terms such as $A^2B^2$ should not be added.
In our defense, a perturbation analysis in the coefficient of such terms indicates that a corrected solution
can be constructed.

It is not clear how to embed such solutions in a UV-complete
gravity theory; a stringy description is not known yet even for the zero temperature case. Such a description would help in finding specific Lifshitz-like field theories with 
gravity duals.
It would be nice to understand the connection (if it exists) between the Lifshitz spacetime and non-Abelian Lifshitz-like gauge theories \cite{Nayak, Horava:2008jf}.

\vspace{1cm}
{\bf Acknowledgements}
We thank Shamit Kachru and Mike Mulligan for correspondence on 
attempts to construct more familiar actions with Lifshitz black hole solutions
and for helpful advice on Fourier transforms.
We also thank Allan Adams, Sean Hartnoll, Alex Maloney, Omid Saremi, T. Senthil 
and Larus Thorlacius for useful discussions and comments.
This work was supported in part by funds provided by the U.S. Department of Energy
(D.O.E.) under cooperative research agreement DE-FG0205ER41360
and by the National Science Foundation under Grant No. NSF PHY05-51164.

\begin{appendix}
\section{Regularizing the action and boundary stress tensor}
In this appendix, we will show that the on-shell action and boundary stress tensor can be rendered finite by making the action well-behaved, \ie\ the action is stationary on-shell 
under an arbitrary normalizable variation of the bulk fields,
and the boundary terms in the action must not break the residual gauge symmetries of the metric. 

We will first find the constraints imposed by finiteness of the free energy, internal energy and pressure on $c_{i}$. 

The free energy of the boundary theory is
$$ -\CF = 
{S_{\text{onshell}} \over \beta} = \half { L_x L_y }\Bigg[{64 c_N - 8c_0 + 16 c_1 +8 c_2 + 6 c_3 + 16 c_4 - 15 c_5\over 32 r_H^4}$$ \be - {32 + 4c_1 + 8 c_0  + 6c_3+ 2c_4 - 5 c_5 \over 16 \epsilon^2 r_H^2} + {24 +2c_3 -c_5 - 8c_N+ 8 c_0\over 8 \epsilon^4}\Bigg] \ee
where $\beta$ is inverse temperature. We must set  $-c_5 + 24 - 8 c_N + 8c_0 + 2c_3 = 0$ and $-c_1-8 - 2c_0 -3/2 c_3 -1/2 c_3 +5/4c_5 = 0$ to get rid of the divergences in the on-shell action. Further, finiteness of the boundary stress tensor and conformal ward identities impose more constraints on the counterterms. 
 

The internal energy of the boundary theory is 
\begin{eqnarray}
{\CE} = -L_x L_y \sqrt{\gamma} T^t_t  = -L_x L_y\Bigg({16 + 8 c_0 - 2 c_3 + 3 c_5 + 8 c_N\over 8 \epsilon^4}  &&
\\  - {
 32 + 8 c_0 + 4 c_1 - 6 c_3 - 2 c_4 + 15 c_5\over 16 r_H^2 \epsilon^2} 
-
{ 8 c_0 - 16 c_1 - 8 c_2 + 6 c_3 + 16 c_4 - 45 c_5 + 64 c_N\over 64 r_H^4}\Bigg) &&
 \end{eqnarray}
Similarly, the expression for pressure is
$$ \CP  = {1\over 2} L_x L_y \sqrt{\gamma} T^{i}_{i}=L_x L_y \sqrt{\gamma} T^{x}_{x} =
\half { L_x L_y }\Bigg[{64 c_N - 8c_0 + 16 c_1 +8 c_2 + 6 c_3 + 16 c_4 - 15 c_5\over 32 r_H^4} $$ \be - {32 + 4c_1 + 8 c_0  + 6c_3+ 2c_4 - 5 c_5 \over 16 \epsilon^2 r_H^2} + {24 +2c_3 -c_5 - 8c_N+ 8 c_0\over 8 \epsilon^4}\Bigg] \ee
Note that $\CF = -\CP$, as expected in the grand canonical ensemble. Hence, the condition for the divergences in pressure to cancel is same as the condition for divergences in the on-shell action to cancel. However, finiteness of energy imposes additional constraints on the counterterms. In the case of Schr\"odinger black hole, 
it is not possible to get rid of the divergence in the energy without the Neumannizing term \cite{Adams:2008wt}. 

The conformal Ward identity for conservation of the dilatation current  requires 
$z\CE = d\CP$, and in our discussion $d=z=2$.  
The residual gauge freedom of the metric is broken if this condition is not satisfied (see \cite{Son:2008ye}). Note that making the boundary stress tensor finite does not ensure this condition.  We must set $c_2 = 7/2$ for the conformal Ward identity to hold. 
After imposing these conditions, we find
\be \label{EPF} \CE  = \CP = -\CF = L_x L_y {15 - 2c_1 -26 c_N \over 16 r_H^4}\ee

In order to have a well-defined variational principle, we must ensure that $\delta S =0$ onshell. We shall now determine the value of $c_1$ using this condition\footnote{We have determined the value of $c_1$ for the case where Dirichlet boundary condition is imposed on the gauge field. However, the method is general and can be used for other boundary conditions as well.}. The variation of the action is 
$$ \delta S  = \int_{\text{bulk}} \text{EOM} +\half  \int_{\text{bdy}} d^3 x \sqrt{\gamma} \Bigg[ T^{\mu}_{\nu} \delta \gamma^{\nu}_{\mu} +   \({(c_N-1)}e^{-2 \Phi} n_\nu F^{\nu \mu} + \( c_3+c_4 \Phi + 2 c_5 A^2\) A^\mu\) \delta{A_{\mu}} +$$ \be \label{varS} {c_N} A^{\mu} \delta\(n^\nu e^{-2\Phi}F_{\nu\mu}\) - \half \(c_1 + 2c_2 \Phi - c_4 A^2 - 4 c_N A^\mu n^\nu F_{\nu \mu} e^{-2\Phi} \) \delta \Phi\Bigg ] \ee
The first term vanishes onshell. Therefore, the boundary terms must also vanish onshell. Let us assume,  for convenience that Dirichlet boundary condition is imposed on the gauge field ($c_N =0$). Prescribing boundary conditions is equivalent to prescribing the coefficient of the non-normalizable mode of the solution. The allowed variations at the boundary fall faster than the non-normalizable part of the solution, \ie,
\be   \begin{matrix} 
       \delta \gamma^{\mu}_{\nu} = \delta \gamma^{\mu}_{\nu(1)}r^2 + \delta \gamma^{\mu}_{\nu(2)}r^4 + \dots \\
      \delta A_{\mu} = r^{-2}\(\delta A_{\mu(1)}r^2 + \delta A_{\mu(2)}r^4 + \dots \) \\
      \delta \Phi = r^2 \delta \Phi_{(1)} + r^4 \delta \Phi_{(2)} + \dots
   \end{matrix}
\ee
Substituting these expressions in (\ref{varS}) and using the conditions on $c_i$ for energy and pressure to be 
finite\footnote{$c_0 = -(17 - c_1)/8$, $c_3 = -5-c_1$, $c_4 = -2c_1$ and $c_5 = -3-c_1$, when $c_N = 0$.}, 
we find
\be \delta S  = \int  d^3 x\(\sqrt{\gamma}T^{\mu}_{\nu} r^2 \delta \gamma^{\nu}_{\mu(1)} + {\CO}\(r^2\)\delta A_{\mu(1)} + \({c_2 - c_1\over r_H^2}\)\(\delta \Phi_1 + \CO\(r^2\)\)\)\ee
Since $\CE$ and $\CP$ are finite, the first term in the integrand vanishes at the boundary. Hence, $c_1 = c_2 = 7/2$ for the variation of the action to vanish on-shell. Using the values of $c_i$ found above in (\ref{EPF}) we get
$$\CE = \CP = -\CF  = {L_x L_y \over 2 r_H^4}$$
After restoring factors of $8 \pi G$, 
$$\CE = \CP = -\CF =  {L_x L_y \over 16 \pi G r_H^4} = - \half T {\del \CF \over \del T} = \half T \CS$$
We have shown that the stress tensor and on-shell action can be regularized by making the action well-behaved,
 \ie\  $\delta S $ must vanish on-shell and the counterterms should not break any residual gauge symmetry. 
\end{appendix}


\begin{thebibliography}{99}

\bibitem{Sachdev:2008ba}
S.~Sachdev and M.~Mueller,
``Quantum Criticality and Black Holes,''
arXiv:0810.3005 [cond-mat.str-el].


\bibitem{Hartnoll:2009sz}
  S.~A.~Hartnoll,
  ``Lectures on holographic methods for condensed matter physics,''
  arXiv:0903.3246 [hep-th].
  

\bibitem{Ardonne:2003wa}
  E.~Ardonne, P.~Fendley and E.~Fradkin,
  ``Topological order and conformal quantum critical points,''
  Annals Phys.\  {\bf 310}, 493 (2004)
  [arXiv:cond-mat/0311466].


\bibitem{Hsu}
B.~Hsu, M.~Mulligan, E.~Fradkin, E.~A.~Kim,
`` Universal entanglement entropy in 2D Quantum critical points''
arXiv:0812.0203 [cond-mat.stat-mech].

\bibitem{Fradkin}
E.~Fradkin,
``Scaling of Entanglement Entropy at 2D Quantum Lifshitz fixed points and topological fluids''
arXiv:0906.1569 [cond-mat.str-el].


%


\bibitem{Park:2008ki}
  D.~S.~Park,
  ``Graviton and Scalar Two-Point Functions in a CDL Background for General
  Dimensions,''
  JHEP {\bf 0906}, 023 (2009)
  [arXiv:0812.3172 [hep-th]].

\bibitem{Senthil}
P.~Ghaemi, A.~Vishwanath, T.~Senthil,
``Finite temperature properties of quantum Lifshitz transitions between valence bond solid phases: An example of `local' quantum criticality,"
Phys.\ Rev.\ {\bf B 72}, 024420 (2005)
arXiv:cond-mat/0412409v1 [cond-mat.str-el].



  
%

%
%
  
\bibitem{KLM}
  S.~Kachru, X.~Liu and M.~Mulligan,
  ``Gravity Duals of Lifshitz-like Fixed Points,''
  Phys.\ Rev.\  D {\bf 78}, 106005 (2008)
  [arXiv:0808.1725 [hep-th]].



\bibitem{Adams:2008zk}
  A.~Adams, A.~Maloney, A.~Sinha and S.~E.~Vazquez,
  ``1/N Effects in Non-Relativistic Gauge-Gravity Duality,''
  JHEP {\bf 0903}, 097 (2009)
  [arXiv:0812.0166 [hep-th]].
  
\bibitem{Taylor:2008tg}
  M.~Taylor,
  arXiv:0812.0530 [hep-th].

\bibitem{Danielsson}
  U.~H.~Danielsson and L.~Thorlacius,
  ``Black holes in asymptotically Lifshitz spacetime,''
  arXiv:0812.5088 [hep-th].
  
  \bibitem{Omid}
N.~Lashkari, A.~Maloney, O.~Saremi, unpublished.

\bibitem{Azeyanagi:2009pr}
  T.~Azeyanagi, W.~Li and T.~Takayanagi,
  ``On String Theory Duals of Lifshitz-like Fixed Points,''
  arXiv:0905.0688 [hep-th].
  
  
\bibitem{taknogo}
W.~Li, T.~Nishioka and T.~Takayanagi,
``Some No-Go Theorems for String Duals of Non-Relativistic Lifshitz-Like   Theories,''
arXiv:0908.0363 [hep-th].

\bibitem{Li:2008dq}
W.~Li, W.~Song and A.~Strominger,
``Chiral Gravity in Three Dimensions,''
JHEP {\bf 0804} (2008) 082
[arXiv:0801.4566 [hep-th]];
D.~Anninos, W.~Li, M.~Padi, W.~Song and A.~Strominger,
``Warped $\mathrm{AdS}_3$ Black Holes,''
JHEP {\bf 0903} (2009) 130
[arXiv:0807.3040 [hep-th]];
A.~Maloney, W.~Song and A.~Strominger,
``Chiral Gravity, Log Gravity and Extremal CFT,''
arXiv:0903.4573 [hep-th].






\bibitem{Son:2008ye}
  D.~T.~Son,
  ``Toward an AdS/cold atoms correspondence: a geometric realization of the
  Schroedinger symmetry,''
  Phys.\ Rev.\  D {\bf 78}, 046003 (2008)
  [arXiv:0804.3972 [hep-th]].
  
\bibitem{Balasubramanian:2008dm}
  K.~Balasubramanian and J.~McGreevy,
  ``Gravity duals for non-relativistic CFTs,''
  Phys.\ Rev.\ Lett.\  {\bf 101}, 061601 (2008)
  [arXiv:0804.4053 [hep-th]].

\bibitem{Herzog:2008wg}
  C.~P.~Herzog, M.~Rangamani and S.~F.~Ross,
  ``Heating up Galilean holography,''
  JHEP {\bf 0811}, 080 (2008)
  [arXiv:0807.1099 [hep-th]].

\bibitem{Adams:2008wt}
  A.~Adams, K.~Balasubramanian and J.~McGreevy,
  ``Hot Spacetimes for Cold Atoms,''
  JHEP {\bf 0811}, 059 (2008)
  [arXiv:0807.1111 [hep-th]].
  
  
\bibitem{Maldacena:2008wh}
  J.~Maldacena, D.~Martelli and Y.~Tachikawa,
 ``Comments on string theory backgrounds with non-relativistic conformal
  symmetry,''
  JHEP {\bf 0810}, 072 (2008)
  [arXiv:0807.1100 [hep-th]].

\bibitem{VBPK}
  V.~Balasubramanian and P. Kraus
 ``A Stress Tensor For Anti-de Sitter Gravity,"
  arXiv:9902121 [hep-th].
  
  
\bibitem{Son:2002sd}
  D.~T.~Son and A.~O.~Starinets,
  ``Minkowski-space correlators in AdS/CFT correspondence: Recipe and
  applications,''
  JHEP {\bf 0209}, 042 (2002)
  [arXiv:hep-th/0205051].
  
  

\bibitem{Nayak}
M.~Freedman, C.~Nayak, K.~Shtengel,
Phys.\ Rev.\ Lett.\ {\bf 94}, 147205 (2005)
[arXiv:cond-mat/0408257]


\bibitem{Horava:2008jf}
  P.~Horava,
  ``Quantum Criticality and Yang-Mills Gauge Theory,''
  arXiv:0811.2217 [hep-th].

%

  

\bibitem{Birmingham:2001dt}
  D.~Birmingham, I.~Sachs and S.~Sen,
  ``Exact results for the BTZ black hole,''
  Int.\ J.\ Mod.\ Phys.\  D {\bf 10}, 833 (2001)
  [arXiv:hep-th/0102155].
  
\bibitem{Peet:1993vb}
  A.~W.~Peet, L.~Susskind and L.~Thorlacius,
  ``Tachyon hair on two-dimensional black holes,''
  Phys.\ Rev.\  D {\bf 48}, 2415 (1993)
  [arXiv:hep-th/9305030].
  
\bibitem{Bertoldi:2009vn}
  R.~B.~Mann,
  JHEP {\bf 0906} (2009) 075
  [arXiv:0905.1136 [hep-th]];
  G.~Bertoldi, B.~A.~Burrington and A.~Peet,
  arXiv:0905.3183 [hep-th],
  arXiv:0907.4755 [hep-th].
  
\bibitem{Ross:2009ar}
S.~F.~Ross and O.~Saremi,
``Holographic Stress Tensor for Non-Relativistic Theories,''
arXiv:0907.1846 [hep-th].


  
\end{thebibliography}
\end{document}